\documentclass[a4paper,twocolumn,11pt,unpublished]{quantumarticle}
\pdfoutput=1

\usepackage{url}

\usepackage{graphicx}

\usepackage[utf8]{inputenc}
\usepackage[english]{babel}
\usepackage[T1]{fontenc}
\usepackage{amsmath}
\usepackage{hyperref}

\usepackage{graphicx}
\usepackage{subfigure}
\usepackage{epstopdf}
\usepackage{graphicx}

\usepackage{amsfonts}
\usepackage{amssymb, color, bm}

\usepackage{currvita}

\usepackage{algorithmicx}
\usepackage[noend]{algpseudocode}
\usepackage{algorithm,setspace}

\usepackage{physics}

\usepackage{multicol}
\usepackage{lipsum}
\usepackage{mwe}

\usepackage{tikz}
\usepackage{lipsum}

\begin{document}

\title{Learning from Radio using Variational Quantum RF Sensing}

%\author{Ivana Nikoloska, \IEEEmembership{Member, IEEE}
\author{Ivana Nikoloska}
%\thanks{I. Nikoloska is with the Signal Processing Systems Group,  Department of Electrical Engineering, Eindhoven University of Technology, Eindhoven, 5612 AP, The Netherlands (e-mail: i.nikoloska@tue.nl). }
\affiliation{Signal Processing Systems Group,  Department of Electrical Engineering, Eindhoven University of Technology, Eindhoven, 5612 AP, The Netherlands}
\email{i.nikoloska@tue.nl}
%\thanks{The work of I. Nikoloska has been supported by Quantum Delta NL, grant number ... }
%}

%\markboth{Journal of \LaTeX\ Class Files, Vol. X, No. X, January 2026}
\maketitle

\begin{abstract}
In modern wireless networks, radio channels serve a dual role. Whilst their primary function is to carry bits of information from a transmitter to a receiver, the intrinsic sensitivity of transmitted signals to the physical structure of the environment makes the channel a powerful source of knowledge about the world. In this paper, we consider an agent that learns about its environment using a quantum sensing probe, optimised using a quantum circuit, which interacts with the radio-frequency (RF) electromagnetic field. We use data obtained from a ray-tracer to train the quantum circuit and learning model and we provide extensive experiments under realistic conditions on a localisation task. We show that using quantum sensors to learn from radio signals can enable intelligent systems that require no channel measurements at deployment, remain sensitive to weak and obstructed RF signals, and can learn about the world despite operating with strictly less information than classical baselines.

\end{abstract}

%\begin{IEEEkeywords}
%wireless communication networks, quantum machine learning, quantum sensing, quantum computing
%\end{IEEEkeywords}

\section{Introduction }
\subsection{Context and Motivation} 
Over the past several decades, five generations of wireless systems have progressively transformed how people and machines are connected, enabling ever-greater capacity, coverage, and service diversity, ranging from analog voice in 1G, to broadband data and massive IoT connectivity in 5G \cite{yu20175g}. As 6G is beginning to take shape, radio channels have become rich media that embed measurable signatures of the physical environment that can be used beyond their role in supporting reliable data transmission \cite{wild20236g}.

In particular, in wireless systems, the combined effects of propagation phenomena such as scattering, diffraction, reflection, and refraction resulting from objects in the environment give rise to multipath whereby multiple copies of the original signal travel along different paths.
The multipath parameters that define the channel state information (CSI) effectively describe how the transmitted signal traversed the environment en route to the receiver \cite{tse2005fundamentals, popovski2020wireless}. Specifically, the amplitude, phase, and path delay resulting from each propagation path capture distinct aspects of the propagation environment. The amplitude of a path encodes information about the attenuation suffered due to distance, shadowing by obstacles, and reflectivity of surfaces, thereby capturing variations in the geometry and material properties of the environment. Similarly, the time delay associated with each path corresponds to the geometric length of that path relative to the line-of-sight component, providing implicit clues about the spatial arrangement of reflectors and scatterers. The phase of a received component, influenced by accumulated propagation delay and any additional phase rotation from reflection, carries fine-grained information valuable for inferring relative positioning within the environment.

\begin{figure*}
    \centering
    \includegraphics[width=0.75\linewidth]{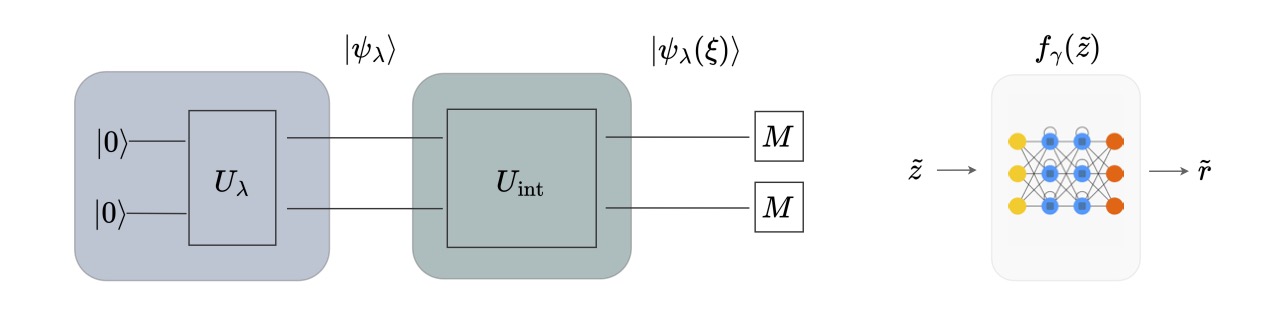}
    \caption{Learning from radio waves via quantum sensing: An agent uses a quantum sensing probe $\ket{\psi_{\lambda}}$ that interacts with the incident RF electromagnetic field $\xi$, modeled as a unitary transformation $U_{\text{int}}$ derived from the rotating wave approximation of the physical interaction. The agent then measures the perturbed state $\ket{\psi_{\lambda}(\xi)}$, producing measurements $\tilde{z}$, and learns to make predictions $\tilde{r}$ using a machine learning model $f_{\gamma}(\tilde{z})$. Both the quantum circuit parameters $\lambda$ and the neural network parameters $\gamma$ are jointly optimised during training. 
    }
    \label{fig:qEM}
\end{figure*}

Recent research has demonstrated that the CSI can be leveraged for environmental sensing and mapping, effectively re-purposing communication signals from carriers of bits to enablers of situational awareness. Techniques such as channel charting use the spatial and temporal structure embedded in the CSI to learn representations of the local radio geometry, enabling tasks such as user localisation or environmental mapping without external positioning references \cite{studer2018channel}. In indoor environments, fine-grained analysis of CSI has been used to extract human motion, showing that even subtle changes in the propagation environment reveal themselves as measurable variations in amplitude and phase across multipath components \cite{zhou2023high}. Moreover, deep learning methods applied to channel characteristics have been shown to identify environmental conditions — for example, differentiating between types of surrounding terrain or weather — by exploiting the sensitivity of CSI to environmental changes \cite{ribouh2022vehicular}.

Because the channel encodes fine-grained spatial and temporal characteristics of the environment, accurately capturing this information requires a highly sensitive sensing apparatus. High sensing precision is essential to resolve small variations in the channel that carry valuable information about the propagation medium. Recently, a parallel trajectory of advances in quantum sensing 
highlights the ability of this technology to detect and measure minute signal variations \cite{degen2017quantum}. Quantum sensors exploit uniquely quantum mechanical effects — such as superposition, entanglement, and coherence — to detect extremely weak signals with sensitivities that can surpass those of classical sensors \cite{helstrom1969quantum}. These devices are inherently sensitive to electromagnetic fields and can be engineered to operate across broad frequency ranges, including the radio‑frequency (RF) bands relevant to wireless communication \cite{simons2021rydberg, cui2025towards}. 

Practically, capturing the benefits of quantum sensing is far from straightforward, since quantum sensors must operate on current hardware that is constrained by noise and physical size. Modern noisy intermediate-scale quantum (NISQ) devices, in particular, suffer from decoherence and sampling errors. Finding near-optimal quantum probe states for a specific sensing task involves searching over extremely high-dimensional spaces, making direct computational approaches largely infeasible. One approach to navigate the complexity of quantum optimisation in the NISQ era is to use variational methods \cite{cerezo2021variational}, which has given rise to frameworks for  variational quantum sensing (VQS) \cite{nolan2021machine,xiao2022parameter}. A VQS scheme uses parameterised quantum circuits to prepare probe states that are adaptively optimised to suit the parameter estimation task. Prior works have demonstrated the potential of VQS focusing on enhancing  estimation precision \cite{meyer2021variational,maclellan2024end}, or improving estimation reliability \cite{nikoloska2025dynamic, nikoloska2025adaptivebayesiansingleshotquantum}.

\subsection{Main Contributions} 
Motivated by these advances, in this paper we investigate agents that use quantum RF sensing probes to learn from the incident RF electromagnetic fields as shown in Fig.~\ref{fig:qEM}. For example, the agent can learn to predict its position, or headings, to navigate unfamiliar terrain.  The main contributions are summarised as follows.

\noindent \textbf{Variational quantum RF sensing:} We first develop a framework for variational quantum RF sensing whereby a quantum sensing probe, optimised using a variational quantum circuit, interacts with the incident RF electromagnetic field. We derive a rotating wave approximation of the interaction from first principles, yielding an efficient training procedure suitable for NISQ hardware. Unlike existing VQS schemes, we are not attempting to solve  the respective estimation problem directly. We are interested in enabling the agent to make predictions with the RF electromagnetic field serving as stimuli.

\noindent \textbf{Learning from the RF electromagnetic field:} The agent uses a machine learning model to process the resulting information, effectively integrating sensing and learning. To train the quantum circuit and learning models, we use data obtained from a ray-tracer, and we do not assume any additional structural knowledge, e.g., the positions of the transmitters, or types of objects/materials in the environment. Once deployed, the proposed scheme does not actually require access to any channel state measurements; rather, the state of the quantum sensor is altered by the incident RF electromagnetic field, which then results in relevant information for the learning model. 

\noindent \textbf{Experimental validation:} Finally, we provide extensive experiments in realistic conditions using a localisation task in which the agent must determine whether it has reached specific target locations in the deployment environment. 
%We choose two distinct targets: one target that has a transmitter nearby and, as a result, strong line-of-sight paths, and a second target target is hidden behind obstacles. 
The results show that the use of variational quantum RF sensing can give rise to intelligent systems that can learn about the world directly from radio waves.

\noindent \textbf{Organisation:} The remainder of the paper is organised as follows. Sec.~II provides some preliminaries on wireless communication channels. In Sec.~III we derive the interaction between the sensor and the RF field and in Sec.~IV we present the proposed scheme for learning from radio waves. In Sec.~V we provide experiments testing the performance of the proposed scheme, including details on the setup, benchmarks, and results. Sec.~VI offers a discussion and concludes the paper.

\section{Wireless Communication Channels}
In a wireless communication system, the fundamental objective is to reliably transfer bits of information between spatially separated endpoints without requiring a physical connection. To this end, a transmitter maps its message into bits of information \(b \in \{0,1\}\) which are to be sent via the wireless channel, grouped into symbols $s_k$. For example, Quadrature Amplitude Modulation (QAM) symbols are generated by mapping groups of $k$ bits to complex values as
\begin{align}
    s_k = |s_k|e^{i\theta_k},
\end{align}
where $\theta_k$ denotes the phase \cite{rappaport2002wireless}. 

The continuous-time baseband signal is then formed by pulse shaping as
\begin{align}
x_{\text{bb}}(t) = \sum_{k} s_{[k]}\, p(t - kT),
\end{align}
where \(p(t)\) is a pulse shaping filter (e.g., raised-cosine) and \(T\) is the symbol period. The baseband signal is then upconverted to a carrier frequency \(f_c\) as
\begin{align}
x(t)= \Re\left\{x_{\text{bb}}(t)\,e^{j2\pi f_c t}\right\}.
\end{align}
The choice of carrier frequency is largely driven by spectrum allocation, propagation physics, and the performance needs of the service. Different generations of wireless technologies operate in distinct ranges of the radio spectrum because of regulatory spectrum allocation and performance trade-offs \cite{tikhomirov2018recommended}. 
%Earlier generations like 2G typically used lower frequency bands, e.g., $850-900$ MHz or $1800-1900$ MHz, which provide wide coverage and good penetration through obstacles but limited peak capacity due to narrow bandwidth. By contrast, 5G New Radio (NR) is designed to use a wider variety of spectrum, including low-band (below $\sim 1$ GHz) for coverage, mid-band ($\sim 2-6$ GHz) for a balance of range and capacity, and even millimeter-wave bands above $\sim 24$ GHz to exploit much larger available bandwidths for very high data rates. 
The frequencies used by wireless communication systems are also referred to as RF.

\begin{figure}
    \centering
    \includegraphics[width=0.9\linewidth]{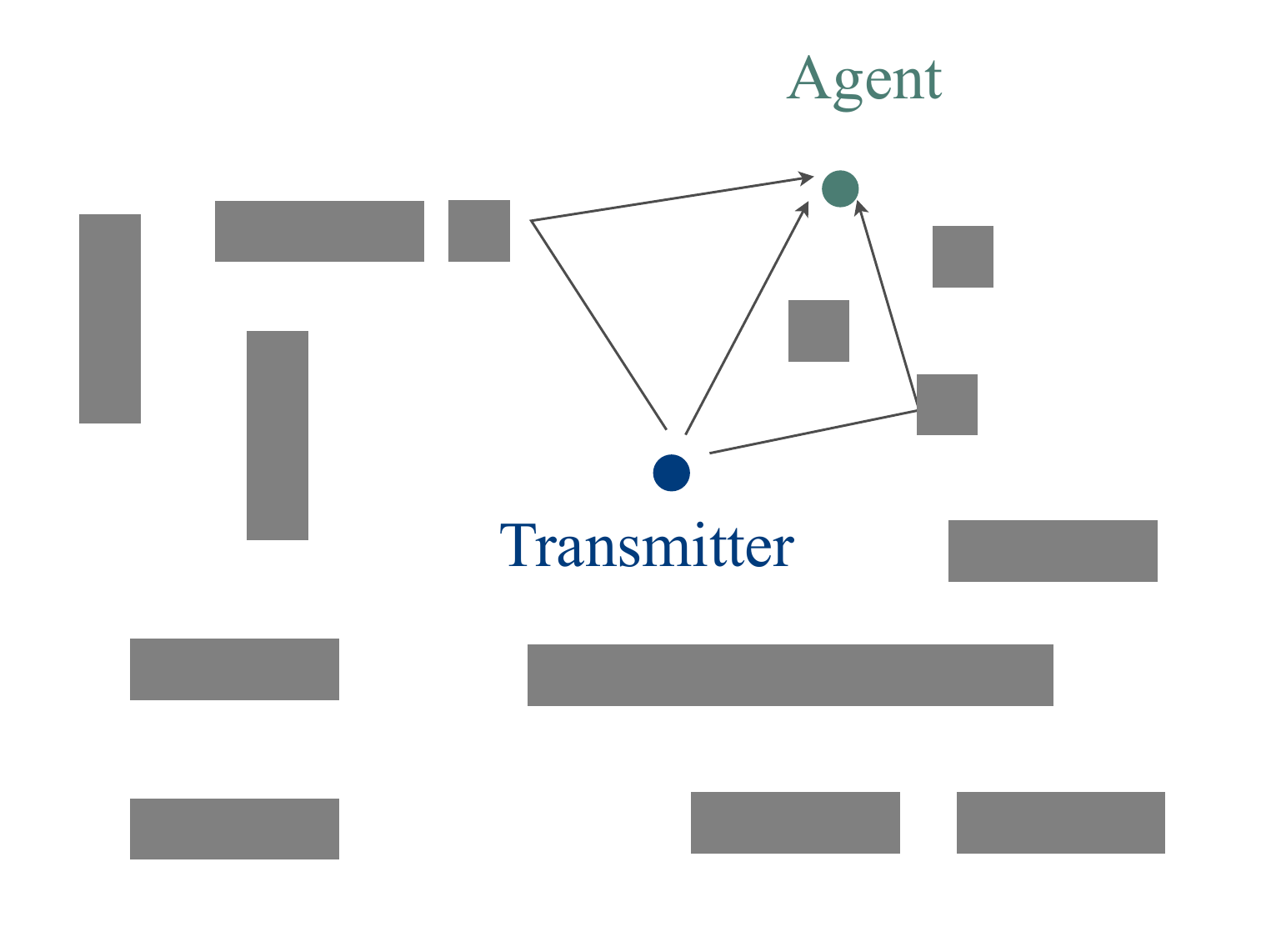}
    \caption{In a realistic environment, many physical components affect the propagation of radio waves. Structures such as buildings, vehicles, and terrain features cause the transmitted signal to undergo reflection, diffraction, and scattering. These result in multiple copies of the original signal traveling along different paths — a phenomenon known as multipath propagation.}
    \label{fig:mtp}
\end{figure}

This real-valued signal then drives an antenna which radiates an electromagnetic wave into space. In particular, the antenna acts as a transducer that converts the high‑frequency electrical signal at its terminals into electromagnetic radiation propagating into space. When the voltage \(x(t)\) drives current through the antenna, this time‑varying current produces time‑varying charge and current distributions on the antenna structure, which in turn generate radiated RF electromagnetic waves (alternatively, radio waves).
These radio waves occupy designated RF bands and propagate through the environment, carrying the encoded information. 

The propagation of radio waves is governed by the principles of electromagnetism and is influenced by both the physical medium and the surrounding environment \cite{tse2005fundamentals, popovski2020wireless}. As a result, at the receiver, the signal is shaped by the propagation environment. 
In free space, radio waves attenuate according to the inverse-square law: as the distance between the transmitter and receiver increases, the signal power decreases rapidly due to spatial spreading. 
%This attenuation, commonly referred to as free-space path loss, forms the baseline for link budget analysis and is a critical component in propagation modeling. 
In realistic environments, however, many additional physical mechanisms affect the propagation of radio waves. Structures such as buildings, vehicles, and terrain features cause the transmitted signal to undergo reflection, diffraction, and scattering. As shown in Fig.~\ref{fig:mtp}, these result in multiple copies of the original signal traveling along different paths — a phenomenon known as multipath propagation. Each multipath component arrives at the receiver with a different delay, amplitude, and phase, leading to constructive and destructive interference that causes rapid fluctuations in the received signal, often described as fading.

The resulting RF electromagnetic field at the receiver is given by \cite{cui2025towards} 
\begin{align}\label{RF_field}
\xi(t)
&=\sum_{k=1}^{K}\sum_{l=1}^{L}
\mu_{eg}^T\,\epsilon_{kl}\,
\sqrt{P_k\rho_{kl}}\,
|s_k|\nonumber\\
&
\times\cos(\omega (t-\tau_{kl})+\phi_{kl}+\theta_k),
\end{align}
where  $k$ indexes transmitters and $l$ indexes propagation paths,  $\mu_{eg}$ is the atomic transition dipole moment, $\epsilon_{kl}$ is the field polarization unit vector, $P_k$ is transmit power, $\tau_{kl}$ is the propagation delay, and $\rho_{kl}$ is the amplitude. Using \eqref{RF_field}, the intended receiver would then attempt to decode the transmitted symbols ${s}_k$ and retrieve the transmitted message.

In this scenario, the delay of each path $\tau_{kl}$ reflects the geometric length of that path and the interactions it has encountered. Indirect paths impose larger delays and the resulting delay spread across all multipath components is itself a measure of the richness and complexity of the channel. 
%This delay profile carries implicit information about the overall structure of the physical environment, capturing the spread in time of the received signal due to environmental complexity. 
Similarly, the amplitude $\rho_{kl}$ of each path depends on the degree of attenuation experienced along that path which arises from path length, absorption by materials, and the number and type of reflections encountered. Stronger amplitudes may indicate a relatively unobstructed path or surfaces with high reflectivity, whereas weaker amplitudes suggest either longer propagation distances, losses due to surface absorption, or more complex scattering. 
Phase shifts arise directly from the path length as $-\omega \tau_{kl}$ and reflection and refraction in the environment which introduce additional phase rotation $\phi_{kl}$; as a result, the phase term of a multipath component contains fine-grained information about the interaction history of that component. 
These parameters form a signature of the propagation channel which reflects the presence, distribution, and properties of any objects and obstacles along the way. We show next how this can be used in order to learn tasks that require situational awareness.

\section{Variational Quantum RF Sensing}
We consider an agent that uses a quantum sensing probe interacting with the incident RF electromagnetic field. The agent then makes predictions based on the accrued information.
The agent does not decode any messages and it does not have any structural information about the environment, including the presence of objects or obstacles, types of material, or the number or positions of transmitters, making a prediction using only the interaction with the RF field.

In particular, the evolution over time of a quantum state $\ket{\psi} = \alpha \ket{g} + \beta \ket{e}$, where $\ket{g}$ and $\ket{e}$ denote the ground and excited states, is governed by Schrödinger's equation as \cite{cui2025towards}
\begin{align}\label{sch}
    i \hbar \frac{\partial \ket{\psi }}{\partial t} = H \ket{\psi },
\end{align}
where
\begin{align}\label{inter}
    H=H_0+H_1(t).
\end{align}
In \eqref{inter}, the free Hamiltonian $H_0$ is given by
\begin{align}
    H_0=\frac{\hbar\omega_{0}}{2}\sigma_{z},
\end{align}
and the time-dependent term $H_1(t)$ is given by
\begin{align}\label{propH}
{H}_1(t)&=
\xi(t)\,\ket{e}\bra{g} +
\xi^*(t)\,\ket{g}\bra{e},
\nonumber\\
\end{align}
with $\xi(t)$ defined in \eqref{RF_field}. The evolution in \eqref{inter} can be transformed using the so-called rotating frame \cite{fleming2010rotating, agarwal1971rotating, zeuch2020exact}. In particular, we first define the rotating frame state
\begin{equation}
\vert \tilde{\psi}\rangle
= U^\dagger(t)\vert \psi\rangle
= e^{i\omega t\,\sigma_z/2}\,\vert \psi \rangle,
\end{equation}
where $\sigma_z$ denotes the Pauli $Z$ operator.
%For a static state, the rotating frame vector $\vert \tilde{\psi}_{\lambda}\rangle$ is a solution to the problem in \eqref{sch} with the Hamiltonian $H= H_0$ and initial state $\ket{\psi_{\lambda}}$.
To transform the Hamiltonian into the rotating frame, we consider the time evolution of a vector in the rotating frame given by \cite{fleming2010rotating}
\begin{align}
\frac{\partial \vert \tilde{\psi}\rangle}{\partial t} 
&= (\partial_t U^\dagger(t))\,\vert \psi\rangle
+ U^\dagger(t)\,\partial_t\vert \psi\rangle\\
&= \Big((\partial_t U^\dagger(t))\,U(t)
- i\,U^\dagger(t)\,H\,U(t)\Big)\,\vert \tilde{\psi}\rangle.
\end{align}
Moreover, using
\begin{equation}
\Xi=\sum_{k=1}^{K}\sum_{l=1}^{L}
\mu_{eg}^T\,\epsilon_{kl}\,
\sqrt{P_k\rho_{kl}}\,
\,|s_k|e^{i(-\omega \tau_{kl} + \varphi_{kl} + \theta_k)},
\end{equation}
we define the Rabi frequency and phase as
\begin{align}
    \Omega = |\Xi|
\end{align}
and
\begin{align}
    \Phi = \text{arg} (\Xi),
\end{align}
respectively. With these, we can write 
\begin{align}\label{propH1}
{H}_1(t)
&\propto \Omega\,\cos(\omega t+\Phi)\,\sigma_x, 
\end{align}
where $\sigma_x$ denotes the Pauli $X$ operator, and we can identify the so-called rotating wave Hamiltonian as
\begin{align}\label{rotH}
&\tilde{H}
= i\,(\partial_t U^\dagger(t))\,U(t)
+ U^\dagger(t)\,H\,U(t)\nonumber\\
&= \hbar\frac{\omega_0-\omega}{2}\,\sigma_z
+ e^{i\omega t\,\sigma_z/2}\,\Omega\,\cos(\omega t+\Phi)\,\sigma_x\,
e^{-i\omega t\,\sigma_z/2}\nonumber\\
&= \hbar\frac{\omega_0-\omega}{2}\,\sigma_z
+ \Omega\,\cos(\omega t+\Phi)\,
\begin{pmatrix}
0 & e^{i\omega t}\\
e^{-i\omega t} & 0
\end{pmatrix}.
\end{align}
Expanding $\cos(\cdot)$ using Euler’s formula 
\begin{equation}
\cos(\omega t+\Phi)
= \frac{1}{2}\Big(e^{(i\omega t+i\Phi)} + e^{-(i\omega t+i\Phi)}\Big)
\end{equation}
and plugging it into \eqref{rotH} results in
\begin{align}
\tilde{H}
&= \hbar\frac{\omega_0-\omega}{2}\,\sigma_z
+ \frac{1}{2}\,\Omega \nonumber\\
&\times
\begin{pmatrix}
0 & e^{-i\Phi}+e^{2i\omega t+i\Phi}\\
e^{-2i\omega t-i\Phi}+e^{i\Phi} & 0
\end{pmatrix}.
\end{align}
The rapidly oscillating terms at \(2\omega\) average out over timescales of interest and can be neglected. Thereby, for the rotating frame approximation we have
\begin{align}
\tilde{H}
&\approx \hbar\frac{\omega_0-\omega}{2}\,\sigma_z
+ \frac{1}{2}\,\Omega\,
\begin{pmatrix}
0 & e^{-i\Phi}\\
\,e^{i\Phi} & 0
\end{pmatrix}
\end{align}
which can be expressed in the Pauli basis as
\begin{equation}\label{effH}
\tilde H
\approx\frac{\hbar\Delta}{2}\sigma_{z}
+\frac{\hbar\Omega}{2}\big(\cos(\Phi)\,\sigma_{x}+\sin(\Phi)\,\sigma_{y}\big),
\end{equation}
where $\Delta = \omega_0 - \omega$ and where $\sigma_y$ denotes the Pauli $Y$ operator.
%The expression in \eqref{effH} can be efficiently implemented on a gate-based NISQ platform. 
Moreover, when \(\Delta=0\), the Hamiltonian in \eqref{effH} simplifies to
\begin{align}\label{effH_1}
\tilde H\approx\frac{\hbar\Omega}{2}\big(\cos(\Phi)\,\sigma_{x}+\sin(\Phi)\,\sigma_{y}\big).
\end{align}
For $N$ particle systems, assuming that the field is uniform and couples independently, the total Hamiltonian becomes 
\begin{align}\label{multiq}
    \tilde H = \sum_{n=1}^N \tilde H^{(n)},
\end{align}
with $\tilde H^{(n)}$ given by \eqref{effH_1}. 

With this, we can efficiently map the interaction to a NISQ device and train the agent as shown in Fig.~\ref{fig:qEM}. In particular, in the variational quantum RF sensing framework, the agent prepares a $N$-qubit quantum sensing probe $\ket{\psi_{\lambda}}$, using a quantum circuit $U_{\lambda}$ parameterised by vector $\lambda$ as
\begin{align}
    \ket{\psi_{\lambda}} = U_{\lambda} \ket{\psi_0}
\end{align}
where $\ket{\psi_0}$ denotes an initial state.
The probe state $\ket{\psi_{\lambda}}$ then interacts with the RF field which is implemented as a unitary transformation 
\begin{equation}\label{int_s}
U_{\text{int}}=\bigotimes_{n=1}^N U^{(n)},
\end{equation}
where $U^{(n)}$ denotes the single-qubit unitary sequence acting on qubit $n$ corresponding to the evolution operator in \eqref{effH_1}. Specifically, as \eqref{effH_1} describes a rotation about an axis in the $XY$-plane at angle $\Phi$ from the $X$-axis, it can be decomposed as a basis-changing $R_{z}(\Phi)$ rotation, followed by an $R_{x}(\Omega t)$ rotation, and the inverse $R_{z}(-\Phi)$, resulting in
\begin{align}
U^{(n)}=R_{z}(\Phi)\,R_{x}(\Omega t)\,R_{z}(-\Phi).
\end{align}
The perturbed state of the quantum sensor after the interaction is then given by
\begin{align}\label{interact}
    \ket{\psi_{\lambda}(\xi)} = U_{\text{int}} \ket{\psi_{\lambda}},
\end{align}
where $U_{\text{int}} \ket{\psi_{\lambda}}$ denotes the application of unitary $U_{\text{int}}$ in \eqref{int_s} to state $\ket{\psi_{\lambda}}$. 
%This approach also allows repeating the unitary application sequence $U_{\lambda}$ and $U_{\text{int}}$ multiple times. In practice, this would be realised by allowing repeated interactions with the RF field. 

\section{Learning from the RF Electromagnetic Field}\label{Sec:LR}
To train the agent, we consider a supervised learning setting where we have access to a training dataset. 
We obtain the training dataset using a simulator, as commonly done in simulation-based inference \cite{cranmer2020frontier}. In general, a simulator builds and maintains a high-fidelity virtual representation of a physical system. In wireless communication, a ray-tracer can track the propagation paths between each transmitter-receiver pair based on the geometry and material information~\cite{hoydis2022sionna, hoydis2023sionna}. In this process, multiple propagation paths are explicitly modeled by considering various propagation effects, such as reflection, scattering, and diffraction. The resulting propagation paths are used to determine the training dataset. 
The use of a ray-tracer for training is a deliberate design choice rooted in the simulation-to-real transfer paradigm \cite{cranmer2020frontier}, whereby the policy developed in simulation is then deployed in the real world without requiring further adaptation. Whilst the ray-tracer does require geometric and material information about the environment during training, this information is only needed once, offline, prior to deployment. Once the variational parameters are learned, the deployed agent requires no channel measurements and no knowledge of the environment — it operates solely through its interaction with the incident RF electromagnetic field. This separation between a knowledge-intensive training phase and a knowledge-free deployment phase is a key design principle of the proposed scheme.
%For each propagation path, and unmodulated carrier, the ray-tracer produces path parameters including complex path gain $a_{kl}$, and propagation delay ${\tau}_{kl}$ where
%\begin{equation}\label{RT}
%|a_{kl}| =
%\mu_{eg}^T\,\epsilon_{kl}\,
%\sqrt{P_k\rho_{kl}},
%\end{equation}
%and 
%\begin{equation}\label{RT1}
%\phi_{kl} =
%-\omega \tau_{kl} + \text{arg}(a_{kl}),
%\end{equation}
%resulting in the sequences $\{\{\{{\rho}_{lk}^m, {\tau}_{kl}^m\}\}_{k=1}^K\}_{l=1}^L\}_{m=1}^M$. 

In particular, for $M$ agent locations in the deployment environment, we obtain simulated RF fields ${\xi}^m (t)$ for $m = 1, ..., M$, which we write as ${\xi}^m$. Together with the corresponding targets ${r}^m$, which depend on the task under consideration, we write the dataset as $\mathcal{D} = \{{\xi}^m, {r}^m\}_{m=1}^M$. We use the simulated fields to determine the interaction unitaries in \eqref{int_s}, and implement the interaction with the RF electromagnetic field by extension.  We note that the framework can also be extended to unsupervised learning, or reinforcement learning settings \cite{goodfellow2016deep}.  
%The ray-tracer can also account for any specific radiation patterns of the transmit antenna, \cite{jiang2023digital}.  

%The agent learns the dataset using a classical machine learning model, or a quantum circuit as discussed next.

%\subsection{Variational Quantum Sensing and Classical Learning}\label{Seq:cq}
As seen in Fig.~\ref{fig:qEM}, to learn the task with a classical machine learning model, the perturbed state $\ket{\psi_{\lambda}({\xi}^m)}$ in \eqref{interact}  is measured which results in 
\begin{align}\label{pred}
    \tilde{z}^m = \langle O \rangle_{{\psi_{\lambda}({\xi}^m)}}.
\end{align}
The result of the measurement $\tilde{z}^m$ is fed into the classical machine learning model $f_{\gamma}(\tilde{z}^m)$ (i.e., a neural network) to obtain a prediction. The goal is to optimise the parameters of the quantum circuit $U(\lambda)$ and the learning model $f_{\gamma}(\tilde{z}^m)$ and minimise a loss function $\mathcal{L}_{\lambda,\gamma}(\cdot)$ between the prediction and the true target. For example, the loss can be defined as the cross-entropy loss in the case of classification as
\begin{align}\label{ce_h}
    &\mathcal{L}_{\lambda, \gamma}({\xi}^m, r^m) = - \log \frac{e^{w^m_{\tilde{r}}}}{\sum_{r'} e^{w^m_{\tilde{r}'}}},
\end{align}
where $w^m_{\tilde{r}}$ denotes the logit corresponding to hypothesis $\tilde{r}^m$ and the sum runs over all hypotheses. In the case of regression, the loss can be defined as the mean-squared error (MSE) loss  whereby
\begin{align}\label{mse_h}
    &\mathcal{L}_{\lambda, \gamma}({\xi}^m, r^m) = (f_{\gamma}(\tilde{z}^m) - r^m)^2.
\end{align}
%or the absolute loss
%\begin{align}
%    &\mathcal{L}_{\lambda, \gamma}({\xi}^m, r^m) = |f_{\gamma}(\tilde{z}^m) - r^m|.
%\end{align}
Formally, we aim to solve
\begin{align}\label{prob_l_cq}
    \underset{\lambda, \gamma}{\text{arg min}} \,\,\, \frac{1}{M}\sum_{m=1}^M \mathcal{L}_{\lambda,\gamma}({\xi}^m, r^m),
\end{align}
where the optimisation is done over the variational parameters $\lambda, \gamma$. This problem is solved via gradient descent whereby the parameters are updated as
\begin{align}\label{grad_phi}
    \lambda \leftarrow \lambda - \eta \frac{\partial \mathcal{L}_{\lambda, \gamma}({\xi}^m, r^m)}{\partial\lambda},
\end{align}
and
\begin{align}\label{grad_w}
    \gamma \leftarrow \gamma - \eta \frac{\partial \mathcal{L}_{\lambda, \gamma}({\xi}^m, r^m)}{\partial \gamma},
\end{align}
where $\eta$ denotes a learning rate. The gradients in \eqref{grad_phi} can be obtained via parameter-shift rules on quantum hardware, or via standard backpropagation on a simulator \cite{mitarai2018quantum, schuld2019evaluating, nikoloska2026machine}. The gradients in \eqref{grad_w} can also be computed via standard backpropagation. Training is done \textit{in silico}, using the ray-tracer data, and, assuming transferability, one could then use the same variational parameters when the RF field is unknown in a real environment. 

 \section{Experiments}
In this section, we provide experimental results to validate the proposed scheme. The code for the experiments will be available on Github.

\subsection{Task}
\begin{figure}
    \centering
    \includegraphics[width=0.9\linewidth]{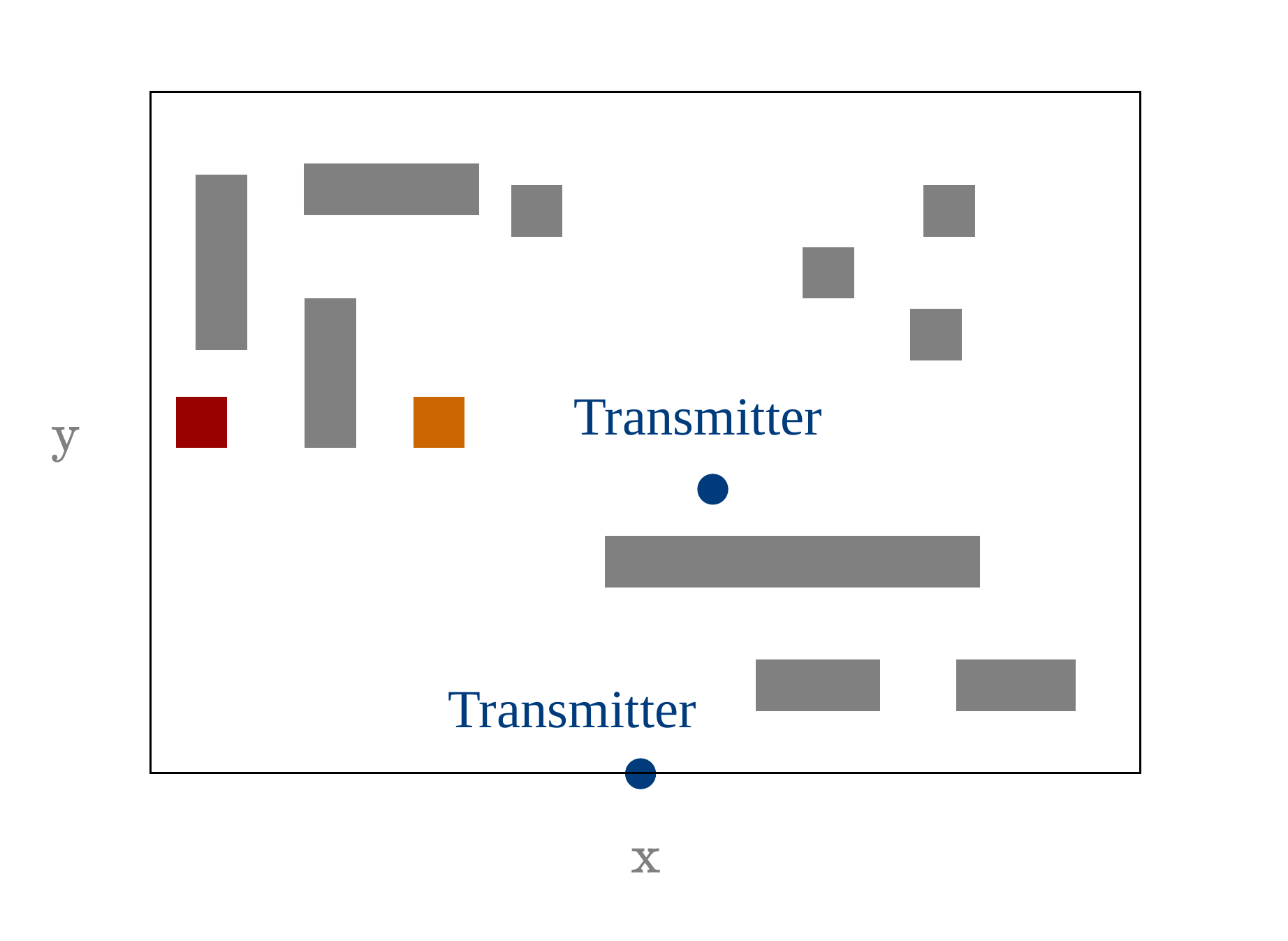}
    \caption{We consider an urban scenario with two transmitters (blue circles) communicating in an urban environment with multiple buildings (grey rectangles). We assume that communication occurs at a frequency $f_c = 2.14$ GHz. The transmitter has single-antenna equipment and the agent's location $[\rm{x}, \rm{y}]$ is generated uniformly at random within the deployment area. We consider two target locations: one which has a transmitter nearby and, as a result, strong line-of-sight path (orange rectangle), and the second which is hidden behind an obstacle (red rectangle).}
    \label{fig:scen}
\end{figure}

We consider the task of localisation, where an agent must determine  whether it reached a predetermined target location \cite{akai2014development, ataka2022magnetic}.
In particular, we consider the network deployment depicted in Fig.~\ref{fig:scen}, in which two transmitters (blue circles) communicate in an urban environment with multiple buildings (grey rectangles). We assume that communication takes place at a frequency $f_c = 2.14$ GHz. The transmitter has single-antenna equipment and the agent's location $[\rm{x}^m, \rm{y}^m]$, where $\rm{x}^m \in [-70, 70]$, $\rm{y}^m \in [-30, 70]$, is generated uniformly at random within the deployment area. We assume unit transmit power. The ray-tracer takes as input the agent's location $[\rm{x}^m, \rm{y}^m]$ and, for an unmodulated carrier, produces the sequence of complex path gains, and propagation delays $\{\{\{a_{kl}^m, {\tau}_{kl}^m\}\}_{k=1}^K\}_{l=1}^L\}_{m=1}^M$, where %$a_{lk}^m = |a_{lk}^m| e^{i {\phi^e}^m_{kl}}$. These are used to determine the interaction unitaries in \eqref{int_s} as
\begin{equation}\label{RT}
a_{kl}^m =
\mu_{eg}^T\,\epsilon_{kl}\,
\sqrt{P_k\rho_{kl}^m} e^{i {\phi}^m_{kl}}.
\end{equation}
%and 
%\begin{equation}\label{RT1}
%\phi_{kl} =
%-\omega \tau_{kl} + \text{arg}(a_{kl}).
%\end{equation}
%We use the $m$-th sequence to determine the interaction unitaries in \eqref{int_s}, and model the precession in the RF electromagnetic field by extension according to \eqref{inter}. 
%For $M$ simulated fields ${\xi}^m$ obtained at different locations in the deployment environment and corresponding targets $r^m$, we write the training dataset as $\mathcal{D} = \{{\xi}^m, r^m\}_{m=1}^M$. 
We use the Sionna ray-tracer \cite{hoydis2022sionna} and we generate the dataset under ground-truth material parameters. Whilst Sionna does not inherently support quantum receivers, using an isotropic ("iso") antenna at the agent location ensures that the ray tracer captures the full incoming field without any hardware-specific filtering. 
%We use $M=2000$ data samples, with $0.8-0.2$ train-test split. 

\begin{figure*}[htbp]
    \centering
    \subfigure{\includegraphics[width=0.32\textwidth]{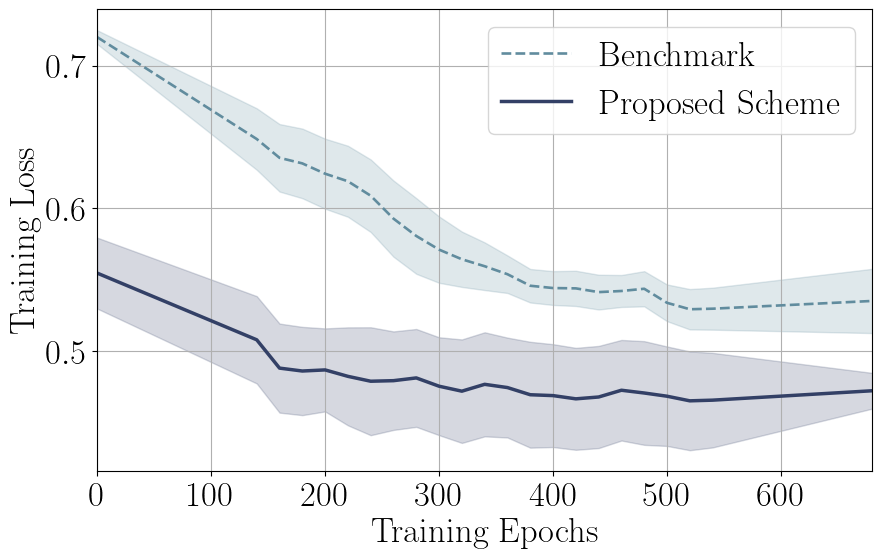}}
    \subfigure{\includegraphics[width=0.32\textwidth]{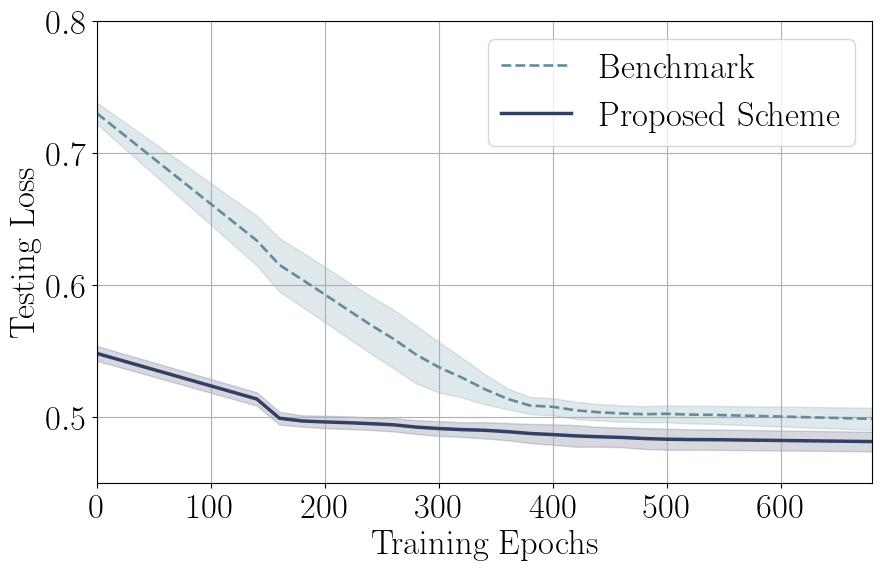}}
    \subfigure{\includegraphics[width=0.32\textwidth]{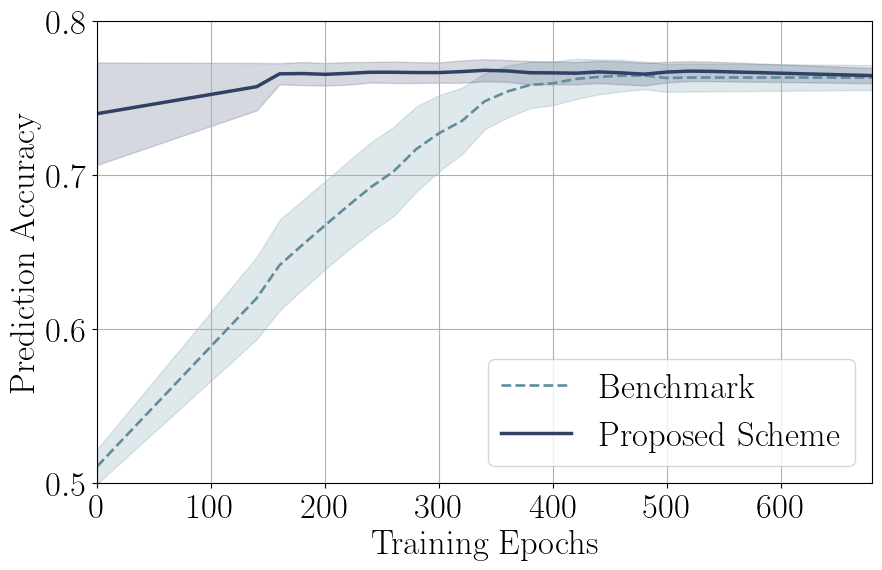}}
    \caption{Accrued training loss (left), testing loss (center), and prediction accuracy (right) over training epochs. We consider the target location shown in Fig.~\ref{fig:scen} (orange rectangle) which has a transmitter nearby and strong line-of-sight path. All results are averaged over three independent trials. Shaded regions represent the variance across trials, which reflects random initialisation and dropout during training.}
    
    \label{fig:entropy_Tr}
\end{figure*}

\begin{figure*}[htbp]
    \centering
    \subfigure{\includegraphics[width=0.32\textwidth]{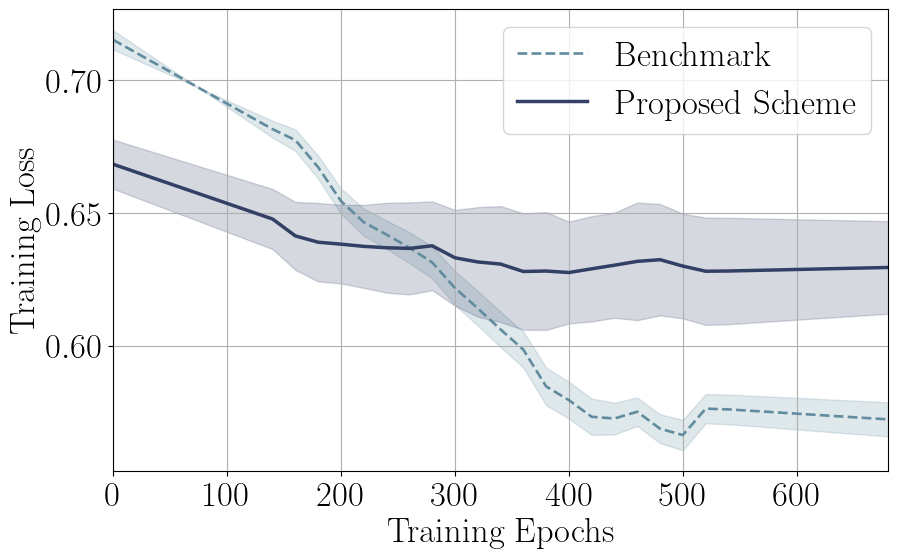}}
    \subfigure{\includegraphics[width=0.32\textwidth]{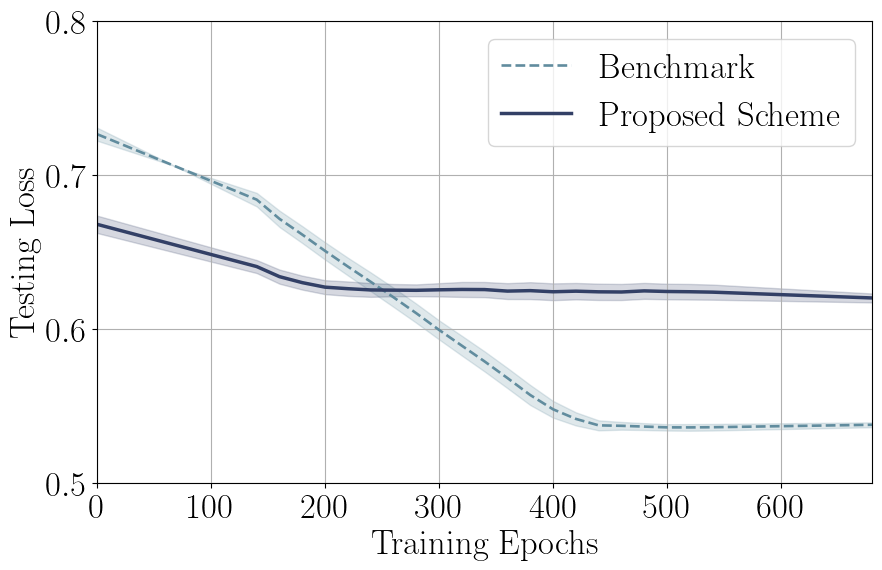}}
    \subfigure{\includegraphics[width=0.32\textwidth]{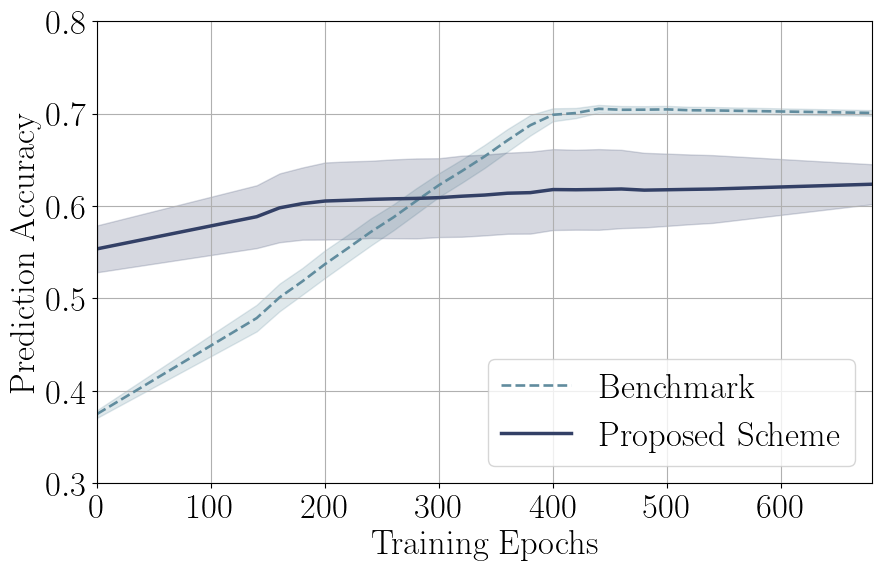}}
    \caption{Accrued training loss (left), testing loss (center), and prediction accuracy (right) over training epochs. We consider the target location shown in Fig.~\ref{fig:scen} (red rectangle) which is hidden behind obstacles. All results are averaged over three independent trials. Shaded regions represent the variance across trials, which reflects random initialisation and dropout during training.}
    
    \label{fig:entropy_Tr1}
\end{figure*}

%The transmitters location ...
The agent's goal is to learn to predict whether it reached the target after the interaction with the RF field. We specifically chose two distinct target locations, shown in Fig.~\ref{fig:scen}, whereby one target (orange rectangle) has a transmitter nearby and, as a result, strong line-of-sight path, and the second target (red rectangle) is hidden behind an obstacle. Each target region is a rectangle of size $10\times10$. Specifically, for the first target location the corresponding target labels are given by
\begin{align}
    r^m =
\begin{cases}
    &0, \,\,\,-30 \leq \rm{x}^m \leq -20, 10 \leq \rm{y}^m \leq 20 \nonumber\\
    &1, \,\,\, \text{otherwise},
\end{cases}
\end{align}
for $m = 1, ..., M$, and for the second target we have
\begin{align}
    r^m =
\begin{cases}
    &0, \,\,\,-70 \leq \rm{x}^m \leq -60, 10 \leq \rm{y}^m \leq 20 \nonumber\\
    &1, \,\,\, \text{otherwise},
\end{cases}
\end{align}
for $m = 1, ..., M$. We note that, if the agent could use the position of the transmitters, this problem could be solved with standard triangulation techniques. Here however, we assume that the agent has no knowledge about the environment and only uses the interaction to make a prediction at deployment.

\subsection{Architecture and Hyperparameters}\label{sec:arch}
We consider an $N=10$-qubit probe state $\ket{\psi_{\lambda}}$ prepared by a parameterised circuit $U_{\lambda}$ comprised of $5$ layers. Each layer consists of two-qubit $R_Z(\lambda)$ gates, followed by single-qubit $R_Z(\lambda)$ and $R_Y(\lambda)$ rotations. The parameters $\lambda$ are not shared between gates and layers. 

We use a compact learning model. In particular, we measure the expected values of Pauli $Z$ observables and feed the results into a fully-connected neural network. The model is comprised of two hidden layers of sizes $128$ and $64$ and ReLU activations. The input layer of size $N$ takes in the measurement outcomes of $\ket{\psi_{\lambda}(\xi^m)}$, and the output layer of size $2$ produces the logits for two classes.  We use dropout with rate $0.2$. The variational parameters are learnt by minimising the cross-entropy loss via \eqref{grad_phi} and \eqref{grad_w} with learning rate $\eta = 0.003$. We use  $M=2000$ data samples, with $0.8-0.2$ train-test split and mini-batches of size $64$. 

%\noindent {\textbf{Small capacity}} For the proposed scheme to be applicable on-device, we also investigate learning models with small capacity. In this case, we allow repeated interactions with the field after each preparation layer, interleaving sensing and computation, by repeating the unitary application sequence $U_{\lambda}$ and $U_{\text{int}}$. We also add a final layer of $R_Y(\lambda)$ rotations following the repeated interactions. The parameters $\lambda$ are not shared between gates. Again, we measure the expected values of Pauli $Z$ observables and feed the results into a final linear layer that takes in the input of size $N$ and produces the logits of size $2$ representing both classes, for a total of ... variational parameters. We use \eqref{grad_phi} and \eqref{grad_w}, with $\eta = 0.02$. 

\subsection{Benchmarks}\label{sec:bench}
As a benchmark, we consider the setting where all multipath parameters are known. This approach is inspired by existing works in the literature that use CSI measurements for localisation tasks, e.g., \cite{studer2018channel, zhou2023high, yu2019indoor, de2020csi}. We specifically choose this scheme since in this case the agent has access to both richer information (all individual multipath  parameters and delay spread by extension), as well as a very powerful model architecture. 

The sequence of complex path gain, and propagation delay $\{\{\{a_{kl}^m, {\tau}_{kl}^m\}\}_{k=1}^K\}_{l=1}^L\}_{m=1}^M$, is used to obtain the inputs to a long short-term memory (LSTM) network \cite{hochreiter1997long} that learns to predict whether the target location has been reached. Specifically, the LSTM model takes as the $m$-th input the phase sequence
\begin{align}
    \phi^m_{kl} = \text{arg} (a^m_{kl}e^{-i \omega \tau^m_{kl}})
\end{align}
for $k = 1,...,K$ and $l = 1, ..., L$. The sequence is time-ordered with the phase $\phi^m_{kl}$ corresponding to the smallest delay $\tau^m_{kl}$ being the first entry, and the phase $\phi^m_{kl}$ corresponding to the largest delay $\tau^m_{kl}$ being the last.
%In particular, the sequence is transformed as
We choose a sequential model to encode an inductive bias into the choice of the architecture, aiming for the best possible performance \cite{bronstein2021geometric}. In addition, LSTM networks naturally handle data samples of different length, as is the case here. The model is comprised of an LSTM layer with hidden size of $64$, followed by a fully connected layer of size $64$ and an output layer of size $2$, for a total of $17,282$ trainable parameters. We use the Adam optimiser with a learning rate of $0.0001$. We use  $M=2000$ data samples, with $0.8-0.2$ train-test split and mini-batches of size $64$.

\subsection{Results}\label{sec:res}

We evaluate the performance of the proposed scheme and the benchmark by examining the accumulated training and testing loss, as well the prediction accuracy, defined as the ratio of correctly classified inputs. The training loss is calculated according to \eqref{ce_h}, reflecting how well each method fits the data it was trained on. The testing loss and the prediction accuracy are evaluated on $400$ previously unseen data samples providing a measure of how well the learned model generalises to new inputs not encountered during training.

Figure~\ref{fig:entropy_Tr} shows the evolution of training and testing losses. The proposed scheme successfully learns the underlying task, as evidenced by the rapid decrease and eventual stabilisation of the training loss. In addition, it is consistent across trials and achieves high prediction accuracy after only a few training epochs, demonstrating that the model does not only memorise the training data, but it generalises quickly to unseen samples. 
We compare the proposed scheme with a sequential model that learns from the complete sequence of multipath parameters. 
%The proposed schemes is seen to converge in fewer steps, with either model architecture.
%This highlights two key advantages: (i) faster minimisation of the loss, and (ii) better generalisation. 
The benchmark converges more slowly and plateaus at a higher loss level. 
%Among the proposed schemes, the moderate-capacity model generalises fastest, whilst the model with the small capacity achieves similar training and testing losses.
Overall, the results in Fig.~\ref{fig:entropy_Tr} demonstrate that learning from electromagnetic fields with variational quantum sensors is both fast and accurate, generalising well to unseen data and outperforming the benchmark in convergence speed.

When the target location is hidden behind obstacles, as seen in Fig.~\ref{fig:entropy_Tr1}, the agent can still learn to solve the task using the proposed approach. Overall, the training and testing losses are higher and the prediction accuracy is lower compared to the first target location. However, the agent still achieves comparable prediction accuracy to the classical benchmark.
We note that the benchmark has access to full channel knowledge. The proposed scheme almost matches a fully-informed, classical model with a suitable architecture and double capacity, even in the harder setting, highlighting the sensitivity of quantum sensing to weaker RF electromagnetic fields.
Unlike the benchmark, the proposed approach does not require any channel measurements once deployed and could carry out its task only by interacting with the RF electromagnetic field.

\section{Discussion and Conclusion}
Quantum sensors could significantly expand what intelligent machines can perceive about their environment, enabling capabilities beyond the reach of classical techniques. In this paper, we investigated the use of variational quantum sensing to learn from the incident RF electromagnetic field in wireless communication networks. We provided extensive experiments in realistic conditions on a localisation task, and we showed that using quantum RF sensing can enable machine intelligence that can effectively utilise the radio channel as a powerful source of knowledge about the world.
%, hinting at new types of machine intelligence that combine precise sensing mechanisms with nascent types of stimuli.

Several directions remain for future work. Practical quantum systems are subject to noise and decoherence, and incorporating realistic noise models into the framework is an important next step. Alternative parameterised circuits, such as symmetry-preserving ansätze, could further improve performance whilst reducing circuit complexity \cite{nikoloska2023time}. Although we did not encounter barren plateau issues in our relatively shallow circuits \cite{mcclean2018barren}, advanced optimisers or regularisation strategies to mitigate this phenomenon are also worth exploring, as are gradient-free training methods. Finally, more suitable architectures for the learning model can also be useful.

\bibliographystyle{IEEEtran}
\bibliography{litdab}

\begin{thebibliography}{10}
\providecommand{\url}[1]{#1}
\csname url@samestyle\endcsname
\providecommand{\newblock}{\relax}
\providecommand{\bibinfo}[2]{#2}
\providecommand{\BIBentrySTDinterwordspacing}{\spaceskip=0pt\relax}
\providecommand{\BIBentryALTinterwordstretchfactor}{4}
\providecommand{\BIBentryALTinterwordspacing}{\spaceskip=\fontdimen2\font plus
\BIBentryALTinterwordstretchfactor\fontdimen3\font minus \fontdimen4\font\relax}
\providecommand{\BIBforeignlanguage}[2]{{%
\expandafter\ifx\csname l@#1\endcsname\relax
\typeout{** WARNING: IEEEtran.bst: No hyphenation pattern has been}%
\typeout{** loaded for the language `#1'. Using the pattern for}%
\typeout{** the default language instead.}%
\else
\language=\csname l@#1\endcsname
\fi
#2}}
\providecommand{\BIBdecl}{\relax}
\BIBdecl

\bibitem{yu20175g}
H.~Yu, H.~Lee, and H.~Jeon, ``What is 5g? emerging 5g mobile services and network requirements,'' \emph{Sustainability}, vol.~9, no.~10, p. 1848, 2017.

\bibitem{wild20236g}
T.~Wild, A.~Grudnitsky, S.~Mandelli, M.~Henninger, J.~Guan, and F.~Schaich, ``6g integrated sensing and communication: From vision to realization,'' in \emph{2023 20th European Radar Conference (EuRAD)}.\hskip 1em plus 0.5em minus 0.4em\relax IEEE, 2023, pp. 355--358.

\bibitem{tse2005fundamentals}
D.~Tse and P.~Viswanath, \emph{Fundamentals of wireless communication}.\hskip 1em plus 0.5em minus 0.4em\relax Cambridge university press, 2005.

\bibitem{popovski2020wireless}
P.~Popovski, \emph{Wireless Connectivity: An Intuitive and Fundamental Guide}.\hskip 1em plus 0.5em minus 0.4em\relax John Wiley \& Sons, 2020.

\bibitem{studer2018channel}
C.~Studer, S.~Medjkouh, E.~Gonulta{\c{s}}, T.~Goldstein, and O.~Tirkkonen, ``Channel charting: Locating users within the radio environment using channel state information,'' \emph{IEEE Access}, vol.~6, pp. 47\,682--47\,698, 2018.

\bibitem{zhou2023high}
H.~Zhou, Y.~Zhang, and M.~Temiz, ``High-resolution indoor sensing using channel state information of wifi networks,'' \emph{Electronics}, vol.~12, no.~18, p. 3931, 2023.

\bibitem{ribouh2022vehicular}
S.~Ribouh, R.~Sadli, Y.~Elhillali, A.~Rivenq, and A.~Hadid, ``Vehicular environment identification based on channel state information and deep learning,'' \emph{Sensors}, vol.~22, no.~22, p. 9018, 2022.

\bibitem{degen2017quantum}
C.~L. Degen, F.~Reinhard, and P.~Cappellaro, ``Quantum sensing,'' \emph{Reviews of modern physics}, vol.~89, no.~3, p. 035002, 2017.

\bibitem{helstrom1969quantum}
C.~W. Helstrom, ``Quantum detection and estimation theory,'' \emph{Journal of statistical physics}, vol.~1, no.~2, pp. 231--252, 1969.

\bibitem{simons2021rydberg}
M.~T. Simons, A.~B. Artusio-Glimpse, A.~K. Robinson, N.~Prajapati, and C.~L. Holloway, ``Rydberg atom-based sensors for radio-frequency electric field metrology, sensing, and communications,'' \emph{Measurement: Sensors}, vol.~18, p. 100273, 2021.

\bibitem{cui2025towards}
M.~Cui, Q.~Zeng, and K.~Huang, ``Towards atomic mimo receivers,'' \emph{IEEE Journal on Selected Areas in Communications}, vol.~43, no.~3, pp. 659--673, 2025.

\bibitem{cerezo2021variational}
M.~Cerezo, A.~Arrasmith, R.~Babbush, S.~C. Benjamin, S.~Endo, K.~Fujii, J.~R. McClean, K.~Mitarai, X.~Yuan, L.~Cincio \emph{et~al.}, ``Variational quantum algorithms,'' \emph{Nature Reviews Physics}, vol.~3, no.~9, pp. 625--644, 2021.

\bibitem{nolan2021machine}
S.~Nolan, A.~Smerzi, and L.~Pezz{\`e}, ``A machine learning approach to bayesian parameter estimation,'' \emph{npj Quantum Information}, vol.~7, no.~1, p. 169, 2021.

\bibitem{xiao2022parameter}
T.~Xiao, J.~Fan, and G.~Zeng, ``Parameter estimation in quantum sensing based on deep reinforcement learning,'' \emph{npj Quantum Information}, vol.~8, no.~1, p.~2, 2022.

\bibitem{meyer2021variational}
J.~J. Meyer, J.~Borregaard, and J.~Eisert, ``A variational toolbox for quantum multi-parameter estimation,'' \emph{npj Quantum Information}, vol.~7, no.~1, p.~89, 2021.

\bibitem{maclellan2024end}
B.~MacLellan, P.~Roztocki, S.~Czischek, and R.~G. Melko, ``End-to-end variational quantum sensing,'' \emph{arXiv preprint arXiv:2403.02394}, 2024.

\bibitem{nikoloska2025dynamic}
I.~Nikoloska, H.~Joudeh, R.~van Sloun, and O.~Simeone, ``Dynamic estimation loss control in variational quantum sensing via online conformal inference,'' \emph{arXiv preprint arXiv:2505.23389}, 2025.

\bibitem{nikoloska2025adaptivebayesiansingleshotquantum}
I.~Nikoloska, R.~Van~Sloun, and O.~Simeone, ``Adaptive bayesian single-shot quantum sensing,'' \emph{arXiv preprint arXiv:2507.16477}, 2025.

\bibitem{rappaport2002wireless}
T.~S. Rappaport, ``Wireless communications--principles and practice, (the book end).'' \emph{Microwave Journal}, vol.~45, no.~12, pp. 128--129, 2002.

\bibitem{tikhomirov2018recommended}
A.~Tikhomirov, E.~Omelyanchuk, and A.~Semenova, ``Recommended 5g frequency bands evaluation,'' in \emph{2018 Systems of Signals Generating and Processing in the Field of on Board Communications}.\hskip 1em plus 0.5em minus 0.4em\relax IEEE, 2018, pp. 1--5.

\bibitem{fleming2010rotating}
C.~Fleming, N.~Cummings, C.~Anastopoulos, and B.-L. Hu, ``The rotating-wave approximation: consistency and applicability from an open quantum system analysis,'' \emph{Journal of Physics A: Mathematical and Theoretical}, vol.~43, no.~40, p. 405304, 2010.

\bibitem{agarwal1971rotating}
G.~Agarwal, ``Rotating-wave approximation and spontaneous emission,'' \emph{Physical Review A}, vol.~4, no.~5, p. 1778, 1971.

\bibitem{zeuch2020exact}
D.~Zeuch, F.~Hassler, J.~J. Slim, and D.~P. DiVincenzo, ``Exact rotating wave approximation,'' \emph{Annals of physics}, vol. 423, p. 168327, 2020.

\bibitem{cranmer2020frontier}
K.~Cranmer, J.~Brehmer, and G.~Louppe, ``The frontier of simulation-based inference,'' \emph{Proceedings of the National Academy of Sciences}, vol. 117, no.~48, pp. 30\,055--30\,062, 2020.

\bibitem{hoydis2022sionna}
J.~Hoydis, S.~Cammerer, F.~A. Aoudia, A.~Vem, N.~Binder, G.~Marcus, and A.~Keller, ``Sionna: An open-source library for next-generation physical layer research,'' \emph{arXiv preprint arXiv:2203.11854}, 2022.

\bibitem{hoydis2023sionna}
J.~Hoydis, F.~A. Aoudia, S.~Cammerer, M.~Nimier-David, N.~Binder, G.~Marcus, and A.~Keller, ``Sionna rt: Differentiable ray tracing for radio propagation modeling,'' in \emph{2023 IEEE Globecom Workshops (GC Wkshps)}.\hskip 1em plus 0.5em minus 0.4em\relax IEEE, 2023, pp. 317--321.

\bibitem{goodfellow2016deep}
I.~Goodfellow, Y.~Bengio, A.~Courville, and Y.~Bengio, \emph{Deep learning}.\hskip 1em plus 0.5em minus 0.4em\relax MIT press Cambridge, 2016, vol.~1, no.~2.

\bibitem{mitarai2018quantum}
K.~Mitarai, M.~Negoro, M.~Kitagawa, and K.~Fujii, ``Quantum circuit learning,'' \emph{Physical Review A}, vol.~98, no.~3, p. 032309, 2018.

\bibitem{schuld2019evaluating}
M.~Schuld, V.~Bergholm, C.~Gogolin, J.~Izaac, and N.~Killoran, ``Evaluating analytic gradients on quantum hardware,'' \emph{Physical Review A}, vol.~99, no.~3, p. 032331, 2019.

\bibitem{nikoloska2026machine}
I.~Nikoloska, ``Machine learning with quantum computers,'' in \emph{Artificial Intelligence and Intelligent Matter: Nanoscience, Soft Matter, Philosophy}.\hskip 1em plus 0.5em minus 0.4em\relax Springer, 2026, pp. 417--434.

\bibitem{akai2014development}
N.~Akai, S.~A. Rahok, K.~Inoue, and K.~Ozaki, ``Development of magnetic navigation method based on distributed control system using magnetic and geometric landmarks,'' \emph{ROBOMECH Journal}, vol.~1, no.~1, p.~21, 2014.

\bibitem{ataka2022magnetic}
A.~Ataka, H.-K. Lam, and K.~Althoefer, ``Magnetic-field-inspired navigation for robots in complex and unknown environments,'' \emph{Frontiers in Robotics and AI}, vol.~9, p. 834177, 2022.

\bibitem{yu2019indoor}
H.~Yu, G.-L. Chen, G.-J. Yu, S.-H. Zhao, B.~Yang, and J.~Liu, ``Indoor passive localisation based on reliable csi extraction,'' \emph{IET Communications}, vol.~13, no.~11, pp. 1633--1642, 2019.

\bibitem{de2020csi}
S.~De~Bast, A.~P. Guevara, and S.~Pollin, ``Csi-based positioning in massive mimo systems using convolutional neural networks,'' in \emph{2020 IEEE 91st Vehicular Technology Conference (VTC2020-Spring)}.\hskip 1em plus 0.5em minus 0.4em\relax IEEE, 2020, pp. 1--5.

\bibitem{hochreiter1997long}
S.~Hochreiter and J.~Schmidhuber, ``Long short-term memory,'' \emph{Neural computation}, vol.~9, no.~8, pp. 1735--1780, 1997.

\bibitem{bronstein2021geometric}
M.~M. Bronstein, J.~Bruna, T.~Cohen, and P.~Veli{\v{c}}kovi{\'c}, ``Geometric deep learning: Grids, groups, graphs, geodesics, and gauges,'' \emph{arXiv preprint arXiv:2104.13478}, 2021.

\bibitem{nikoloska2023time}
I.~Nikoloska, O.~Simeone, L.~Banchi, and P.~Veli{\v{c}}kovi{\'c}, ``Time-warping invariant quantum recurrent neural networks via quantum-classical adaptive gating,'' \emph{Machine Learning: Science and Technology}, 2023.

\bibitem{mcclean2018barren}
J.~R. McClean, S.~Boixo, V.~N. Smelyanskiy, R.~Babbush, and H.~Neven, ``Barren plateaus in quantum neural network training landscapes,'' \emph{Nature communications}, vol.~9, no.~1, p. 4812, 2018.

\end{thebibliography}

\end{document}